\documentclass[%
 reprint,
 amsmath,amssymb,
 aps,unsortedaddress]{revtex4-1}
\usepackage{graphicx}
\usepackage{dcolumn}
\usepackage{bm}
\begin{document}

\title{Partial filled Landau Level at even denominator, a vortex metal with Berry phase }

\author{Yizhi You}
\affiliation{Princeton Center for Theoretical Science, Princeton University, Princeton, New Jersey 08544,USA}

\date{\today}
\begin{abstract}
We develop a vortex metal theory for partial filled Landau Level at $\nu=\frac{1}{2n}$, whose ground state contains a composite Fermi surface(FS) formed by the vortex of electrons. In the projected Landau Level limit, the composite Fermi surface contains $\frac{-\pi}{n}$ Berry phase. Such fractional Berry phase is a consequence of LL projection which produces the GMP guiding center algebra and embellishes an anomalous velocity to the equation of motion for the vortex metal. Further, we investigate a particle-hole symmetric bilayer system with $\nu_1=\frac{1}{2n}$ and $\nu_2=1-\frac{1}{2n}$ at each layer, and demonstrate that the $\frac{-\pi}{n}$ Berry phase on the composite Fermi surface leads to the suppression of $2k_f$ back-scattering between the PH partner bilayer, which could be a smoking gun to detect the fractional Berry phase. We also mention various instabilities and competing orders in such bilayer system including a $Z_{4n}$ topological order phase driven by quantum criticality. 

\end{abstract}

\maketitle

\section{Introduction and motivation}

Landau Fermi liquid mechanism reveals the qualitative structure and universal behavior of interacting Fermi surface at low temperature. However, in low dimension system, there exists a rich class of metallic states beyond the Landau Fermi liquid paradigm. In $1d$, the Tomonaga-Luttinger liquid, whose low energy effective theory can be bosonized, exhibits exotic transport behavior and spin-charge separation in low energy spectrum. In $2d$, the non-Fermi liquid, whose ground state still contains a Fermi surface with decoherent quasiparticle, is another explicit example where the Fermi liquid mechanism breaks down. Candidates of $2d$ non-Fermi liquid\cite{Metlitski-2015b,watanabe2014criterion,senthil2008critical,ruhman2014ferromagnetic,bahri2015stable,you2016nematic,Nayak-1994,oganesyan2001quantum} include the partial filled Landau Level at filling $\frac{1}{2n}$; the strange metal regime near the quantum critical points and/or inside the symmetry breaking phase; the U(1) spin liquids in organic materials. Most of these non-Fermi liquids were explored and observed in experiment, which display unconventional transport properties and lack of well-defined spectral peak. In addition, recent study on SYK models\cite{hartnoll2016holographic,jian2017solvable,berkooz2017higher,song2017strongly} in higher dimension introduces an exact solvable Hamiltonian for non-Fermi liquid. In the large N limit, one could utilize the out-of-time-ordered\cite{blake2017thermal} correlator to characterize the decoherent nature of non-Fermi liquid.

Interacting 2d electrons in the presence of strong magnetic field exhibit rich phase diagram and exotic phenomenon\cite{willett1987observation}.  Beyond the incompressible quantum Hall insulator, there also exists a class of compressible metallic state when electrons are at even filling factor $\nu=\frac{1}{2n}$\cite{willett1987observation,halperin1993theory,Pasquier-1998}.  
Motivated by the idea of composite Fermion(CF) with flux attachment\cite{Jain1989,Lopez-1991}, Halperin, Lee, and Read(HLR)\cite{halperin1993theory} initially developed a theoretical framework for partial filled Landau level with metallic behavior. When each fermion is attached with $2n$ flux, the composite fermion forms a Fermi surface with strong and nonlocal interaction mediated by dynamical gauge fields. The gauge boson is overdamped by the gapless Fermi surface, and the fermion encounters inelastic scattering mediate by gauge boson. Hence, the composite Fermi surface becomes chaotic and decoherent at low $T$ and the quasiparticles acquire finite lifetime\cite{Polchinski-1993,Stern-1999,Nayak-1994,murthy2016nu,Mross-2010}.

The HLR theory provides a simple but primitive framework to clarify the existence of a stable Fermi surface in $\nu=\frac{1}{2n}$ filled Landau Levels(LL). Such composite Fermi surface with non-Fermi liquid nature contains unconventional transport properties. The existence of `composite Fermi surface' with finite wave vector could be probed in experiment by measuring the Weiss oscillation and static dielectric response\cite{du1994drastic}. One can also subtract the Fermi wave vector of the composite Fermi surface\cite{geraedts2016half,balram2015luttinger,shao2015entanglement,mishmash2016entanglement} in numerical simulations by scaling the entanglement entropy with a leading order of $L\ln(L)$ or by figuring out the singularity of Lindhard function, which mostly agrees with the HLR predication.

In the large magnetic field limit where the inter LL gap is much larger than any other interaction scale, one could ignore the effect from LL mixing and project the Hilbert space to the lowest LL in IR theory. In such limit, the half-filled Landau level contains an exact Particle-Hole(PH) symmetry. In addition, for LL bilayer with filling $\nu_1=\frac{1}{2n}$ and $\nu_2=1-\frac{1}{2n}$ at each layer, the system also contains an explicit PH symmetry(up to layer switching). However, based on the previous flux attachment picture in HLR theory, all these phases contain a composite Fermi surface with finite Fermi wave vector. While a Fermi surface indicates non-zero chemical potential, it is paradoxical to get a `survival Fermi surface' in the PH symmetric limit\cite{Kivelson-1997,balram2016nature,levin2016particle,Barkeshli-2015,Lee-2007,Levin-2007,Girvin-1984,wang2016particle,cheung2016weiss,murthy2016nu,lee1999unsettled}.

Recently, Son proposed a new composite Dirac liquid theory for half-filled LL system\cite{Son-2015}, whose composite Fermi surface is charge neutral and contains a $\pi$ Berry phase. In terms of duality mapping, the composite Fermi surface is formed by the vortex of the electron, and perceives the PH symmetry in a way akin to time reversal.
Afterwards, a group of pioneers\cite{seiberg2016duality,wang2016composite,mross2016explicit,wang2016half,Kachru-2015,Metlitski-2015,mulligan2016particle,potter2016realizing,mulligan2016emergent,levin2016particle,balram2016nature} develops similar composite Dirac liquid theory in a microscopic point of view to verify this theory.  

Among these approaches, we get an intuition that the vortex nature of the composite Fermi surface, as well as the $\pi$ Berry phase, is a consequence of LL projection which changes the local Hilbert space and leads to the guiding center algebra with non-commutative geometry\cite{Girvin-1986}.  Motivated by these pioneer study, we expect the partial filled Landau Level at filling $\nu=\frac{1}{2n}$, whose ground state forms a composite Fermi surface, shall contain similar features including a $\frac{\pi}{n}$ Berry phase, as well as the `vortex nature' of the quasiparticle near the Fermi surface, which perceives the PH symmetry in terms of time reversal.

In this paper, we explore the partial-filled Landau Level at even denominator, whose ground state exhibits non-Fermi liquid behavior\cite{wang2016composite}. We start with the partial filled Landau level at $\nu=\frac{1}{2n}$, as well as its PH partner at $\nu=1-\frac{1}{2n}$. These PH partner pairs contain composite Fermi surfaces of the same size, and the quasiparticle on the Fermi surface is formed by vortex of the original electrons. In such vortex metal, the PH symmetry acts on the composite Fermi surface in the form of time-reversal, and this makes the composite Fermi surface survive in the lowest Landau Level limit. We would demonstrate that the guiding center algebra from Landau level projection, creates an anomalous velocity for the motion of composite Fermion and hence generates a $\frac{-\pi}{n}$ ($2\pi+\frac{\pi}{n}$) Berry phase for $\nu=\frac{1}{2n}$ ($\nu=1-\frac{1}{2n}$) LL. Such fractional Berry phase could be verified by the back-scattering between PH partner states whose $2k_f$ singularity is suppressed.  Further, we investigate the various instabilities and competing orders in quantum Hall bilayer at $\nu_1=\frac{1}{2n}$ and $\nu_2=1-\frac{1}{2n}$ filling, and propose an exotic symmetry invariant gapped phase with $Z_{4n}$ topological order which covers a quantum critical region. This result agrees with some recent numerical study\cite{zhuzheng} on half-filled quantum Hall bilayers with unconventional intermediate states.

\section{Overview of the composite fermion theory in half-filled Landau Levels}
For interacting electrons confined in a $2d$ quantum well in the presence of magnetic field at $ \nu=\frac{1}{2}$ filling fraction, the many-body system forms a compressible metallic state with exotic transport behavior. In the long-established HLR approach, the composite fermion, formed by electron attached with two flux quanta, does not perceive the external magnetic field and finally forms a composite Fermi surface at charge neutrality. 
 \begin{align} 
&\mathcal{L}=\Psi^{\dagger,cf}(iD_0+\frac{1}{2m}D^2_{i})\Psi^{cf}+\frac{1}{8\pi}(a-A)\wedge d(a-A)\nonumber\\
&+\frac{1}{g} (da)^2\nonumber\\
& D_{\mu}=\partial_{\mu}+ia_{\mu} 
\end{align}
The composite fermi surface couples with a dynamical gauge theory `$a$', which emerges from the flux attachment procedure. The Chern-Simons term indicates the constraint between flux density and composite fermion density. The Maxwell term describes electron interaction in the density-density channel. The dynamical gauge field creates strong and nonlocal interaction between the composite fermion. The Fermi surface survives under gauge fluctuation, while the quasiparticle near the Fermi surface becomes decoherent and dies away with finite lifetime. Different from the usual Fermi liquid theory where the Fermion would always encounter with pairing instability at low temperature, the pairing channel in our composite fermion theory is strongly suppressed by gauge fluctuation and hence the Fermi surface is stable at zero temperature.

While the HLR theory clarifies the ground state of the half-filled Landau Level(LL) problem, there seems to be a self-contradiction if we go to the lowest Landau level limit. When the electron interaction strength is far below the Landau Level gap, the IR theory merely cares about the electron in the lowest Landau Level. By projecting the physical Hilbert space into the lowest Landau Level, the electron at half filling fraction contains a particle-hole(PH) symmetry.  For a usual PH symmetric theory, one does not expect any survival Fermi surface. As a Fermi surface with finite size contains a non-zero chemical potential, the usual PH operation, which transforms $\Psi^{\dagger} \rightarrow \Psi $, gives a negative chemical potential and hence the theory is not compatible with PH symmetry. Then how does the composite Fermi surface here survive in the lowest LL with PH symmetry? 

This question was under debate for decades until Son\cite{Son-2015} proposed the Dirac description of Half-filled LL via a duality argument. 
 \begin{align} 
&\mathcal{L}=\Psi^{\dagger}_{cf}\gamma_{\mu}(\partial_{\mu}+ia_{\mu} )\Psi_{cf}-\frac{1}{4\pi}a\wedge d A+\frac{1}{8\pi}A\wedge d A\nonumber\\
&+\frac{1}{g} (da)^2
\end{align}
The composite Fermi surface in Son's theory is described by a Dirac fermion at finite chemical potential and hence carries a $\pi$ Berry phase. The Fermi surface still couples with a dynamical U(1) gauge field which finally turns the system into a non-Fermi liquid but the U(1) gauge theory does not have a Chern-Simons term(which breaks PH symmetry).
The PH symmetry here is anti-unitary and acts on the composite Fermi surface in a similar way as time reversal, 
\begin{align} 
& \Psi_{cf} \rightarrow i\sigma_y \Psi_{cf} ,\nonumber\\
& a_0 \rightarrow a_0,\nonumber\\
& a_i \rightarrow -a_i
\end{align}

Concurrently, Wang and Senthil\cite{wang2016composite} proposed a vortex metal theory for half-filled Landau Level via a wave function argument. 

The trial wave function of half-filled LL before LL projection could be written in terms of,
\begin{align} 
\Psi(z_i,z_j......)=\prod_{i<j} (z_i-z_j)~det[e^{i(\bar{k}_i z_i+k_i\bar{z}_i)/2}] (z_i-z_j)
\label{hlrwave}
\end{align}
The $det$ term indicates the composite fermions form a Fermi surface while $(z_i-z_j)^2$ indicates each composite fermion is a combination of electron and $2$ vortices(correlation holes). The composite fermion is at charge neutrality so each vortex(correlation hole) contains charge $-e/2$. The self-statistics of the vortex is $\pi$, which could be retained from the Chern-Simons term.

Before LL projection, the fermion bound with two vortices(correlation holes) form a composite Fermi surface at charge neutrality.  Once we project the Hilbert space into the lowest LL, the GMP algebra\cite{Girvin-1986} of the guiding center coordinate requires the wave-function to be holomorphic so $\bar{z}$ is  replaced with $\partial_z$. Replace this algebra into Eq [\ref{hlrwave}], there appears a translation operator $e^{i \frac {\epsilon^{ij} k_i l_B^2 \partial_j}{2} }$ which
shifts one vortex away from the composite fermion center in a direction orthogonal to the Fermi momentum of the composite fermion\cite{wang2016half}. 

The shifted CF contains a dipole configuration and each dipole-end carries charge $\pm e/2$. Consequently, the vortex-fermion bound state forms a charge dipole and it carries momentum perpendicular to the dipole. This is similar to the spin-orbital coupling in Dirac fermions where the spin and momentum are locked\cite{wang2016half}. In addition, when we go around the Fermi surface, the fermion's momentum angle winds around $2\pi$ and so is the dipole. As the dipole's self-rotation accumulates a $\pi$ Berry phase, the Fermi surface also carries $\pi$ berry phase which exactly matches the theory of Dirac Fermi surface. The PH symmetry rotates the dipole by 180 degrees so one can express the symmetry operator as $\mathcal{K} i\sigma_y$ which rotates the Dirac spinor. The mass term of the Dirac fermion, which breaks PH symmetry is absent here. In addition, as the theory contains a Fermi surface, the only IR theory one could manifest is the composite Fermi surface with $\pi$ Berry phase while the structure far below the Fermi surface is not presented in the IR limit. Hence, there is no guarantee that the fermion far below the Fermi surface is relativistic and the Dirac point is not presented in such theory.

Such wave function picture evidently reveals the origin of `Dirac spinor' in a non-relativistic half-filled LL system. The dipole of the composite fermion, emerge from LL projection plays the role as Dirac spinor and the dipole-momentum locking is responsible for the $\pi$ Berry phase on the Dirac Fermi surface. In addition, such microscopic wave function indicates that the PH symmetry acts on the CF in terms of time-reversal, which compatibly explains the survival of Fermi surface in the presence of PH symmetry.  

Similar dipole argument could be extended to whatever partial filled LL with even denominator $\nu=\frac{1}{2n}$\cite{wang2016composite}. The composite fermion, consists of electron and $2n$ flux quanta, forms a stable Fermi surface. LL projection moves $2n-1$ vortex away from the electron in a direction perpendicular to the Fermi momentum. Such dipole momentum locking ensures a nonzero Berry phase on the Fermi surface. Since the statistical phase acquired by the dipole self-rotation is $\frac{-\pi}{n}$(or $2\pi-\frac{-\pi}{n}$ up to a gauge transformation), the composite Fermi surface also contains $\frac{-\pi}{n}$ Berry phase.

However, the wave function approach in Ref.\cite{wang2016composite} could be subtle as they assume LL projection merely shift one vortex away from the composite fermion center and this is the key process to obtain the correct Berry phase in dipole-momentum locking argument. (If LL projection shift both vortices, the self-rotation of the dipole would accumulate $2\pi$ Berry phase). However, recent numerical simulation\cite{wang2017lattice} indicates such wave function choice would have higher energy. In addition, once we tune the microscopic detail of the system, the wave function would be decorated with more complicated structure and the dipole argument fails.

\section{emergent Berry curvature and anomalous velocity}

As the Berry curvature on the Fermi surface is a consequence of LL projection, the Berry phase should be universal and insensitive to any microscopic details as long as the particle-hole symmetry is preserved. Hereafter, we would demonstrate the existence of the Berry phase via more general argument regardless of any microscopic detail or assumption.  The emergence of a fractional Berry phase of the composite Fermi surface is a consequence of LL projection while the PH symmetry only enhance the robustness of the Berry phase. The GMP algebra and guiding center coordinate, which reflects the key essence of LL projection should be enough to derive the Berry phase in partial filled Landau Level at $\nu=\frac{1}{2n}$.  


Before we proceed, we first revisit the composite Fermi surface in HLR theory, which could be regarded as a Galilean invariant Fermi surface formed by composite particles. Take two quasiparticles on the Fermi surface with momentum $k$ and $k'$, if there exist a phase factor between these two states which cannot be gauged away, there must be a non-vanishing Berry connection between the two states and this contributes to the Berry connection of the Fermi surface. In the composite Fermi surface problem, we choose quasiparticle $\psi (\vec{k})$ with momentum $\vec{k}$ near the composite Fermi surface and generate another quasiparticle $\psi (\vec{k'})$ with momentum $\vec{k'}$ ($k,k' \sim |k_f| $) by, 
\begin{align} 
\psi (\vec{k'})=e^{i \theta_{k-k'} (k_x y-k_y x)}\psi (\vec{k})
\label{rotation}
\end{align} 
$e^{i \theta_{k-k'} (k_x y-k_y x)}$ is the rotation operator in momentum space which changes the Fermi momentum angle by $\theta_{k-k'}$. 
\begin{figure}[h]
\centering
  \includegraphics[width=0.7\linewidth]{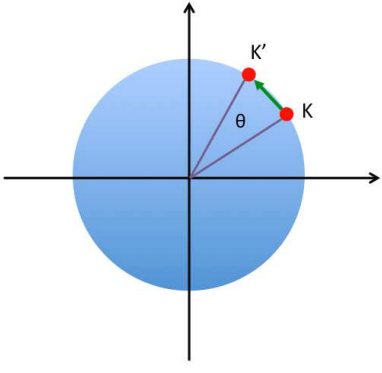}
  \caption{two quasiparticles with momentum $k$ and $k'$ on the Composite Fermi surface}
  \label{fs}
\end{figure}

The Berry connection of the composite Fermi surface without LL projection is trivial and could be gauge away,
\begin{align} 
&\mathcal{A}_{k_i}(k')=    \nonumber\\
&\langle \psi (\vec{k})| e^{-i \theta_{k-k'} (k_x y-k_y x)} ~ \partial_{k_i} ~e^{i \theta_{k-k'} (k_x y-k_y x)}|\psi (\vec{k})\rangle=0\nonumber\\
\end{align}

When we perform LL projection to the composite Fermi surface, the quasiparticle near the Fermi surface is still labeled by the same momentum $k$. Hence one can take Eq [\ref{rotation}] and apply LL projection operator on both sides of the equation.

\begin{align} 
P_{LLL}\psi (\vec{k'})=P_{LLL} e^{i \theta_{k-k'} (k_x y-k_y x)} P_{LLL} \psi (\vec{k})
\label{rotationLLL}
\end{align} 

Once we perform LL projection, the operator $x,y$ in the rotation generator should be replaced by guiding center operator $R_x, R_y$. The motion of the composite Fermion in the lowest LL is characterized by the guiding coordinate $R_i$ with non-commutative geometry.
\begin{align} 
&P_{LLL}r_iP_{LLL}=R_i  \nonumber\\
&R_i=r_i-\epsilon^{ij} \Pi_j/B, ~[R_x,R_y]=-i(l_B)^2
\end{align} 
Hence, the rotation generator in momentum space becomes
\begin{align} 
&P_{LLL}e^{i \theta (k_x R_y-k_y R_x)}  P_{LLL}=  \nonumber\\
&e^{i \theta (k_x R_y-k_y R_x)} =e^{i\theta  (k_x y-k_y x)/2} ~ e^{-i \theta (k_x^2+k_y^2)/B}
\end{align} 
In IR theory, we only focus on the quasiparticle near the composite Fermi surface at wave vector $k_f$, 
\begin{align} 
&e^{-i \theta_{k-k'} (k_x^2+k_y^2)/B} P_{LLL} \psi (\vec{k})\nonumber\\
&=e^{-i \theta_{k-k'} (k_f^2)/B}P_{LLL} \psi (\vec{k})=e^{-i \theta_{k-k'} \nu} P_{LLL} \psi (\vec{k})
\end{align} 
Hence, the wave function of quasiparticle on the FS with momentum $k'$ and $k$ has the relation,
\begin{align} 
P_{LLL} \psi (\vec{k'})=e^{i \theta_{k-k'}(k_x y-k_y x)/2}  e^{-i \theta_{k-k'} \nu} P_{LLL} \psi (k)
\end{align} 
We do not have an explicit form for the wave function $P_{LLL} \psi (k)$ as it could be dependable to some microscopic details of the many-body system. However, It is transparent that the LL projection creates a nontrivial phase factor $e^{-i \theta_{k-k'} \nu}$ between two states on the Fermi surface with momentum $k, k'$. Such phase factor emerges due to the LL projection and guiding center algebra. The phase factor is universal without any microscopic assumption and cannot be gauged away. (Meanwhile, the phase $e^{i \theta_{k-k'} \nu (k_x y-k_y x)/2} $ does not contribute to the Berry curvature.)

When we have filling fraction $\nu=\frac{1}{2n}$, the Berry phase of the Fermi surface is,
\begin{align} 
&\int_{FS} \epsilon^{ij} \partial_{k_j}\mathcal{A}_{k_i} dk_x dk_y =    \nonumber\\
&= -2\pi \nu=\frac{-\pi}{n}
\end{align} 
The Berry phase of the composite Fermi surface is $-2\pi \nu$. 
For $\nu=1/2$, the Berry phase would be $-\pi$(or $\pi$ up to a gauge transformation). Such emergent Berry phase due to LL projection is universal and is merely a consequence of the non-commutative algebra $[R_x,R_y]=-i(l_B)^2$ for the guiding center coordinates, which controls the motion of the composite fermion near the Fermi surface.

For the conventional Fermi surface with non-vanishing Berry connection, the Berry phase comes from the spin-orbital coupling so the Fermi momentum is locked with the spin(or pseudo-spin) orientation near the Fermi surface. The Berry curvature contributes an anomalous velocity to the equation of motion for the electron and hence generates a non-quantized Hall response.

In a metallic system, the semiclassical equation of motion of an electron can be written as\cite{xiao2010berry,haldane2004berry},
\begin{align} 
&\frac{d k_a}{dt}=E_a+\epsilon^{abc}B^c \frac{d x_b}{dt}
\nonumber\\
&\frac{d x_a}{dt}=\frac{\partial \epsilon(k)}{\partial k_a}+F^{ab}(k) \frac{dk_b}{dt}
\label{niu}
\end{align} 
Here we take charge and light speed to unity. $E,B$ are the electromagnetic field and $F^{ab}(k)$ is the antisymmetric Berry curvature at momentum $k$. $\frac{\partial \epsilon(k)}{\partial k_a}$ is the group velocity of the fermion wave packet. $F^{ab}(k) \frac{dk_b}{dt}$ is the anomalous velocity part which suggests the wave packet of the fermion is self-rotating when moving along the group velocity direction. The Berry curvature $F^{ab}(k)$ performs as the Lorenz Force in momentum space.
Combining two equations in Eq[\ref{niu}] one obtains the charge Hall response $J_b=\int_{FS} ~dk_x dk_y F^{ab}(k)~ E_a$. The total Berry curvature $\phi=\int_{FS} ~ dk_x dk_y F^{ab}(k)$ of the filled states determines the non-quantized Hall conductivity contributed from Fermi surface.

For our composite Fermi surface at $\nu=\frac{1}{2n}$ LL, the Fermi surface is charge neutral and it does not perceive the external magnetic field.  Before LL projection, the Fermi surface has no Berry phase so we choose a specific Berry connection by taking $\mathcal{A}_i (k)=0$ for all filled momentum. The semiclassical equation of motion for the composite fermion before LL projection is,
\begin{align} 
&\frac{d k_a}{dt}=e_a
\nonumber\\
&\frac{d x_a}{dt}=\frac{\partial \epsilon(k)}{\partial k_a}
\end{align} 
Here $e_a$ is the electric field generated by the dynamical gauge field $a$ (which refers to the electron current).  The composite Fermion without LL projection does not contain anomalous velocity.

\begin{figure}[h]
\centering
  \includegraphics[width=0.8\linewidth]{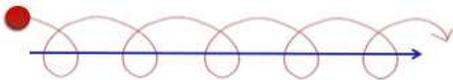}
  \caption{Motion of the composite Fermion in the Lowest LL. The blue line labels the trajectory of the guiding center coordinate, which involves a group velocity, together with an anomalous velocity due to the background of cyclotron motion(red spiral line).}
  \label{gmp}
\end{figure}

Apply the LL projection, the motion of the composite fermion should be labeled by guiding center coordinate $R_a$. Take the definition of $R_i=r_i-\epsilon^{ij} \Pi_j/B$ and insert them to the equation of motion,
\begin{align} 
&\frac{d R_a}{dt}=\frac{d x_a}{2 dt}-\frac{\epsilon^{ab}}{B}\frac{d k_b}{dt}=\frac{\partial \epsilon(k)}{2\partial k_a}-\frac{\epsilon^{ab}}{B}\frac{d k_b}{dt},\nonumber\\
&\frac{d k_a}{dt}=e_a
\label{aaa}
\end{align} 
Even the original Galilean invariant composite Fermi surface does not contain any Berry curvature induced anomalous velocity, once we project the Hilbert space into the lowest LL, the motion of the composite Fermion on the Fermi surface is controlled by the guiding center coordinate $R_i$, which involves the moving of a particle in the background of cyclotron motions as Fig. \ref{gmp}. Hence, the `wave packet of the composite fermion' emerges an anomalous velocity due to the cyclotron motion of the lowest Landau Level.

\begin{align} 
&\phi=-\int_{FS} dk_x dk_y \frac{1}{B}=-(2\pi)^2 \rho^{cf}/B=-2\pi \nu
\label{bbb}
\end{align} 
Hence, for partial filled LL at $\nu=\frac{1}{2n}$ with a metallic ground state, the Berry phase of the composite Fermi surface is $\frac{-\pi}{n}$. Such anomalous velocity on the Fermi surface would create anomalous Hall conductivity $\frac{-e^2}{2n h}$
and generate a Chern-Simons term for the dynamical gauge field $\frac{-\nu}{4\pi} a \wedge da$. For half-filled LL, the anomalous Hall conductivity generated by the Fermi surface cancels the Chern-Simons term from flux attachment, and the theory finally becomes PH invariant.

\section{Vortex metal with a Berry Phase}
\subsection{vortex metal theory for $\nu=\frac{1}{2n}$}
\label{vm}

In this section, we refine the composite Fermion approach for $\nu=\frac{1}{2n}$ Landau Level from a Chern-Simons point of view in the lowest Landau Level limit. As we had demonstrated in our previous content, LL projection creates anomalous velocity and generates Berry phase for the composite Fermi surface. We would further elaborate that the corresponding composite Fermion theory becomes a vortex metal and the Chern-Simons term from flux attachment is canceled due to anomalous velocity.

To operate the Chern-Simons description for $\nu=\frac{1}{2n}$ filled LL, we attach $2n$ correlation holes to the electron. The composite Fermion forms a Fermi surface coupled to a dynamical U(1) gauge theory.
\begin{align} 
&\mathcal{L}=\Psi^{\dagger,cf}(iD_0+\frac{1}{2m} D^2_i)\Psi^{cf}+\frac{1}{8n\pi} (a-A)\wedge d(a-A)\nonumber\\
&+\frac{1}{g}(da)^2\nonumber\\
&D_i =i\partial_i+a_i
\end{align} 
The Chern-Simons gauge field $a$ indicates the constraint between gauge flux and charge density. The Maxwell term $\frac{1}{g}(da)^2$ performs the electron density-density interaction.

In the zero loop level, the Chern-Simons gauge flux is locked with the fermion density and hence cancels the external magnetic flux. The constraint between the composite fermion and gauge fluctuation reads,
\begin{align} 
-\epsilon^{ij}\frac{q_i a_j(q)}{4n\pi}=\rho^e(q)=\sum_{a}e^{iqr^a}
\end{align} 
$\rho^e(q)$ is the electron density and $r^a$ is the coordinate of $a$-th electron before LL projection.

After LL projection, the equation of motion for composite fermion is labeled by guiding center coordinate. From now on, we replace the coordinate operator with $R_i=r_i-\epsilon^{ij}\Pi_j/B$. The constraint between Chern-Simons flux and fermion density thereby becomes,
\begin{align} 
-\epsilon^{ij}\frac{q_i a_j(q)}{4n\pi}=\rho^e(q)=\sum_{a}e^{iq R^a}
\end{align} 
Hence, we can write a modified Chern-Simons theory where the composite Fermion's motion is labeled in terms of $R$,
\begin{align} 
&\mathcal{L}=\Psi^{\dagger, cf}(R)(iD_0+\frac{1}{2m} D^2_i)\Psi^{cf}(R)\nonumber\\
&+\frac{1}{g}(da)^2+\frac{1}{8\pi} (a-A)\wedge d(a-A)\nonumber\\
&D_i =i\partial_i+a_i(R)
\end{align} 
The composite fermion sits at the guiding center coordinate $R$. The theory becomes nonlocal due to the non-commutative geometry of the guiding center algebra. In addition, the gauge theory shall contain a cubic Moyal product term $a_{\mu} \star a_{\nu} \star a_{\rho}$ term\cite{Bigatti-2000} which reflects the non-local feature of the gauge field. However, as this cubic term vanishes at long wave-length limit, we do not include them in our theory.

As we had demonstrated in previous section in Eq [\ref{aaa},\ref{bbb}], the motion of the composite fermion labeled by guiding center coordinate contains anomalous velocity and non-zero Berry curvature. Such anomalous velocity is a consequence of LL projection, which force the composite fermion to move in the background of cyclotron motion of the lowest LL.

\begin{align} 
&\frac{d R_a}{dt}=\frac{d x_a}{2 dt}-\frac{\epsilon^{ab}}{B}\frac{d k_b}{dt}=\frac{\partial \epsilon(k)}{2\partial k_a}-\frac{\epsilon^{ab}}{B}\frac{d k_b}{dt},\nonumber\\
&\frac{d k_a}{dt}=e_a
\end{align} 

The Fermi surface thereby generates a Berry phase $\frac{-\pi}{n}$ together with the additional Chern-Simons term $\frac{-1}{8n\pi}a \wedge da$. This cancels the Chern-Simons term in the original flux attachment theory and the new composite fermion theory with projective LL is,
 \begin{align} 
&\mathcal{L}=\Psi^{\dagger,cf}(iD_0+\frac{1}{2m'}D^2_{i})\Psi^{cf}-\frac{1}{4n\pi}A da+\frac{1}{8n\pi}A dA \nonumber\\
& D_i=\partial_i+ia_i 
\label{a1}
\end{align}
The new theory in the projective LL limit contains a Fermi surface coupling with a dynamical gauge field $a$. The original Chern-Simons term for $a$ is canceled the by anomalous Hall conductivity of the Fermi surface. The Fermi surface contains a Berry phase $-\pi/n$(which is not apparent in Eq [\ref{a1}]). $m'$ is the effective mass of the composite Fermion near the Fermi surface renormalized by Landau parameter $F_1$. We would like to emphasize that Eq [\ref{a1}], as the low energy effective theory for a composite Fermi surface, only involves the fermions near the Fermi surface at $k\sim k_f$. The composite Fermion dispersion far below the Fermi surface is not presented in this theory and is not essentially quadratic.

 Apply the saddle point solution for $a$, we have
 \begin{align} 
&\rho^{cf}=\frac{B}{(2n)2\pi}, \rho^e=-\frac{b}{4n\pi}+\frac{B}{4n\pi}
\end{align}
This suggests the $4n\pi$ gauge flux of $a$ in Eq [\ref{a1}] refers to the electron density while the composite fermion is the vortex of the physical electron. The composite fermion density is bound to the external magnetic field.

The Berry phase of the composite Fermi surface is non-quantized and could flow away to whatever value from $0$ to $2\pi$ under perturbation. In half-filled LL, the $\pi$ Berry phase is robust only if there is an explicit particle-hole symmetry which constrains the Berry phase to be a $Z_2$ invariant. For other metallic states at $\nu=\frac{1}{2n}$ filling, there is no such symmetry to stabilize the Berry phase. However, if we have a particle-hole partner bilayer with $\nu_1=\frac{1}{2n}$ and $\nu_2=1-\frac{1}{2n}$ on each layer, the system contains a particle-hole (up to a layer switching) symmetry. This new symmetry constraint could protect the overall Berry phase of the bilayer metallic system. We would elaborate the symmetry protected Berry phase in the rest part of this paper.

\subsection{vortex metal theory for $\nu=1-\frac{1}{2n}$}
In this part, we consider the PH partner of the $\nu=\frac{1}{2n}$ LL, namely, the partial filled LL at $\nu=1-\frac{1}{2n}$. One could view such states as holes at filling $\nu=\frac{1}{2n}$ in the background of the filled LL.  
Before LL projection, one could write down a Chern-Simons gauge theory by attaching $-2n$ flux quanta to the hole,

\begin{align} 
&\mathcal{L}=\Psi'^{\dagger,cf}(iD_0+\frac{1}{2m}D^2_{i})\Psi'^{cf}-\frac{1}{8n\pi}(a+A)d(a+A)\nonumber\\
&+\frac{1}{4\pi}A dA \nonumber\\
& D_i=\partial_i+ia_i 
\label{hole}
\end{align}
The first two terms imply each hole is attached with $2n$ vortices and such composite particle $\Psi'^{cf}$ forms a Fermi surface. The last term is the Hall conductivity of the background IQHE which fills the lowest LL.

Projecting the theory to the lowest LL, one could duplicate the similar argument in Section \ref{vm}. After LL projection, the composite particle $\Psi'^{cf}$ gains anomalous velocity. The Fermi surface contains a Berry phase of $\pi/n$ and produces the anomalous Hall response $\frac{1}{8n\pi} a \wedge da$. The anomalous Hall conductivity would cancel the Chern-Simons term in Eq [\ref{hole}] due to flux attachment and the consequential theory is,
\begin{align} 
&\mathcal{L}=\Psi'^{\dagger,cf}(iD_0+\frac{1}{2m}D^2_{i})\Psi'^{cf}-\frac{1}{4n\pi} A d a
+\frac{2n-1}{8n\pi}A dA  \nonumber\\
& D_i=\partial_i+ia_i 
\label{newhole}
\end{align}
The composite fermi surface contains a $\pi/n$ Berry phase and share the same Fermi momentum as that in $\nu=\frac{1}{2n}$ filled LL. However, the vacuum of this `composite fermi surface' is the $n=1$ LL which already carries a $2\pi$ Berry phase due to the chirality of IQH state. If we choose the empty LL as the vacuum, the Berry phase of the composite Fermi surface for $\nu=1-\frac{1}{2n} $ would be $2\pi+\pi/n$.

\begin{figure}[h]
\centering
  \includegraphics[width=0.8\linewidth]{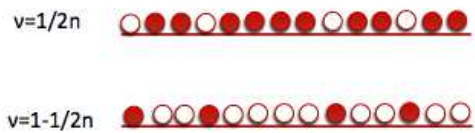}
  \caption{two-layered system with filling fraction $\nu_1=\frac{1}{2n}$ and $\nu_2=1-\frac{1}{2n}$}
  \label{layer1}
\end{figure}

The composite hole liquid at $\nu=1-\frac{1}{2n}$ could be regarded as the PH conjugate of the composite particle liquid at $\nu=\frac{1}{2n}$. If we have a bilayer LL as Fig \ref{layer1} with filling fraction $\nu_1=\frac{1}{2n}$ and $\nu_2=1-\frac{1}{2n}$ on each layer, the system contains PH symmetry(up to a layer switching operation). The two composite fermi surfaces have the same size with Berry phase $-\pi/n$  and $2\pi+\pi/n$. The totally Berry phase $2\pi$ contributed by the bilayer metal is protected by PH symmetry. Although a $2\pi$ Berry phase on a single Fermi surface is trivial, the bilayer system with two Fermi surfaces at identical size has a Berry phase ambiguity to $4\pi$. Hence, the $2\pi$ Berry phase in the bilayer system is still nontrivial and each Fermi surface on the top/bottom layer only contribute a fraction of them.

\subsection{the role of particle-hole symmetry}
While we had explicitly demonstrated that the LL projection creates a Berry phase for the Fermi surface, the role of PH symmetry, and how it acts on the composite fermion level remain unsolved. It is clear that the physical electron is changed into a hole under PH symmetry. However, the composite fermion $\Psi^{\dagger,cf}$ , which consists of both physical electrons operator and correlation vortex projected into the lowest LL, does not necessarily change into composite hole $\Psi^{cf}$ under PH.

In order to investigate the role of PH symmetry for composite fermion, one could check the minimal coupling between composite fermion current and U(1) gauge potential after PH transformation.

For a physical electron current $J_e$ who couples with the EM field as $\vec{J_e} \vec{A}$, after PH transformation, the hole current couples with the EM field as $-\vec{J_h} \vec{A}$. This indicates the PH symmetry switch the creation operator to the annihilation operator and hence carries opposite U(1) charge.

As we had demonstrated in our previous section, the composite fermion current $\vec{J}_{cf}$ merely couples with the dynamical U(1) gauge field.  
 \begin{align} 
&\mathcal{L}=\vec{J}_{cf} \vec{a}-\frac{1}{4n\pi}A da+\frac{1}{8n\pi}A dA 
\label{particle}
\end{align}
PH symmetry takes $(a_0,a_x,a_y)$ to $(a_0,-a_x,-a_y)$.The term $\frac{1}{4n\pi} a\wedge dA$ indicates a $4n\pi$ flux of $a$ is an electron. Hence $\frac{1}{4\pi}  da$ is the physical electron current and the composite fermion could be regarded as the `vortex' of the physical electron.

After PH symmetry transformation from Eq [\ref{a1}] to Eq [\ref{hole}], the theory of composite hole liquid is,
\begin{align} 
&\mathcal{L}=\vec{J'} _{cf}\vec{a} -\frac{1}{4n\pi} A da
+\frac{2n-1}{8n\pi}A dA
\end{align} 
$\vec{J'}_{cf}$ is the composite fermion current after PH transformation which minimal couples with the dynamical gauge field $a$. The term $\frac{1}{4n\pi} a\wedge dA$ indicates $\frac{1}{4n\pi}  da$ is still the electron current. As the composite fermion current after PH transformation still carries the same gauge charge, the `composite fermion' is still a vortex of the electron and PH symmetry transformation does not change composite particle to its antiparticle. Instead, it acts as a $\mathcal{CT}$ symmetry which turns the Berry phase from $-\pi/n$ to $\pi/n$.  As a result, we can conclude that the PH symmetry acting on CF as $\mathcal{CT}$ symmetry. This is straightforward if we regard the CF as the vortex of the physical electron. In particle vortex duality\cite{seiberg2016duality,Metlitski-2015,mross2016explicit}, a particle becomes an anti-particle under PH operation, while a vortex becomes its time reversal partner under PH. The vortex nature of the composite Fermi liquid in $\nu=\frac{1}{2n}$ or $\nu=1-\frac{1}{2n}$ filled LL is crucial as it takes PH symmetry into time-reversal symmetry for the composite Fermi surface, which makes the symmetry compatible with the IR theory in the lowest LL limit.

\subsection{transport signature of the vortex metal}
In Landau Fermi liquid, there is a universal relation between thermal and charge transport $\kappa= \sigma  L_0 T$, know as the Wiedemann-Franz law. Such linear relation simply implies both charge and heat carrier are contributed from the electrons near the Fermi surface. 

For a non-Fermi liquid system, such Wiedemann-Franz paradigm fails and the universal relation between charge and heat transport is absent in most non-Fermi liquids. However, for partial filled Landau Level with a composite Fermi surface, the Fermi surface is formed by vortex of the electron at charge neutrality. Such vortex metal is the heat carrier and hence produce thermal transport $\kappa$ directly. However, as the vortex metal merely couple with the dynamical gauge field $a$, they first bring about the polarization tensor for gauge field $a$ and the dynamics of gauge current which couples with the electromagnetic field finally contributes to the charge response.
The relation between $\kappa$ and $\sigma_{a}$(conductivity of the composite Fermi surface with respect to $a$) is
\begin{align} 
&\kappa= \sigma_a  L_0 T
\end{align} 
Meanwhile, after integrating out the dynamical gauge field, one can readily figure out that $\sigma_{a}$ is with inverse  relation of the charge conductivity $\sigma \sim 1/\sigma_{a}$. Hence, the `Wiedemann-Franz' for vortex metal is
\begin{align} 
&\kappa \sim \frac{1}{\sigma}  L_0 T
\end{align} 
The charge and thermal transport coefficient have inverse relation at fixed temperature\cite{wang2016composite}. This inverse relation is a consequence of the vortex nature for the composite Fermi surface and could be a smoking gun to probe and verify the existence of vortex metal in experiment.

\section{PH partner Landau level bilayer}

\subsection{Robustness of Berry phase under PH symmetry}
Although we had demonstrated the existence of Berry phase due to LL projection, the Berry phase in metallic system(partial filled band) is not quantized and hence can be perturbed to whatever value unless there is some symmetry constraint.
In this section, we study the bilayer Landau level system with $\nu_1=\frac{1}{2n}$ and $\nu_2=1-\frac{1}{2n}$ on each layer. We define a new particle-hole symmetry which involves PH and layer switching operation. Such bilayer system is particle-hole invariant and contains two composite Fermi surface with identical size. The PH symmetry protects the overall Berry phase of $2\pi$ contributed from the two Fermi surfaces on each layer.

Imagine we apply the particle-hole symmetry(together with layer switching) operation $\mathcal{CT}$ to the composite Fermion. 
\begin{align} 
&\mathcal{CT} : \Psi^{cf}_{1,k} \rightarrow e^{i f(k)} \Psi^{cf}_{2,-k} ;\Psi^{cf}_{2,k} \rightarrow e^{i f'(k)} \Psi^{cf}_{1,-k}
\label{abc}
\end{align}
$\Psi^{cf}_{i,k} $ is the composite Fermion with momentum $k$ at $i$ layer. As we had established in our previous section, $\Psi^{cf}_{i,k} $ is the vortex of the physical electron who perceives the PH symmetry as time-reversal.
$f'(k)$ and $f(k)$ are arbitrary phase factors as a function of $k$. Since PH symmetry is projective and anti-unitary, such phase factor cannot be gauged away. Acting PH symmetry twice, we have,
\begin{align} 
&(\mathcal{CT})^2 \Psi^{cf}_{1,k} = \mathcal{CT} e^{i f(k)} \Psi^{cf}_{2,-k} =e^{-i f(k)}e^{i f'(-k)}  \Psi^{cf}_{1,k}
\end{align}
For the bilayer system with $\nu_1=\frac{1}{2n}$ and $\nu_2=1-\frac{1}{2n}$ filling, the whole system has a filling fraction $1/2$. If we act the $\mathcal{CT} $ operator to the many-body system twice, the wave function would gain an overall phase factor $(-1)^{N}$(N is the number of particles). This indicates the system contains Kramers doublet when filling with odd number electrons and each particle gains a phase factor $e^{i \pi}$ when acting PH twice(see Appendix for detailed proof). Therefore, the phase factor in Eq [\ref{abc}] cannot be gauged away,
\begin{align} 
&(\mathcal{CT})^2 \Psi^{cf}_{1,k} =-\Psi^{cf}_{1,k} \rightarrow~e^{-i f(k)}e^{i f'(-k)}=-1
\label{relation}
\end{align}
This relation is merely a consequence of the Kramers degeneracy for PH symmetry in projected LL which does not depend on any microscopic detail. To imply the constraint $~e^{-i f(k)}e^{i f'(-k)}=-1$, we take $f(k)=f'(k)=\theta_k$. Then if the composite Fermion $\Psi^{cf}_{1}$ on the first layer with $\nu=\frac{1}{2n}$ has a Berry phase $\alpha$, the composite Fermion $\Psi^{cf}_{2}$ on the second layer at $\nu=1-\frac{1}{2n}$ should contain a Berry phase $2\pi-\alpha$. Hence the overall Berry phase contributed by the bilayer system is fixed as $2\pi$ and is protected by PH symmetry. As the system contains two independent Fermi surfaces, their Berry curvature can only be gauge away by any integer of $4\pi$ and the $2\pi$ Berry phase here is still nontrivial.

\subsection{$2k_f$ scattering between layers}
In half-filled LL, the $\pi$ Berry phase could be verified by the absence of $2k_f$ back-scattering in the presence of disorder\cite{geraedts2016half}. For a Dirac Fermi surface, the back-scattering near Fermi surface $\Psi^{\dagger, cf}_k \Psi^{cf}_{-k}$ is PH odd. Hence, any PH even operator should have a suppression for singularity at $2k_f$ as the PH odd back-scattering procedure does not contribute at leading order. 
In bilayer system with $\nu_1=\frac{1}{2n}$ and  $\nu_2=1-\frac{1}{2n}$ on each layer, the fractional Berry phase of the composite fermi surface on each layer could be verified in a similar way.

Consider the $2k_f$ back-scattering process between layers,
\begin{align} 
\mathcal{CT}  \Psi^{cf,\dagger}_{1,k} \Psi^{cf}_{2,-k} &=e^{-i f(k)}e^{i f'(-k)} \Psi^{\dagger, cf}_{2,-k} \Psi^{cf}_{1,k}
\end{align}
Take advantage of the relation we obtain in Eq [\ref{relation}] where $e^{-i f(k)}e^{i f'(-k)}=-1$, one could prove that such interlayer back-scattering procedure is PH odd. 
\begin{align} 
\mathcal{CT}  \Psi^{cf,\dagger}_{1,k} \Psi^{cf}_{2,-k} =- \Psi^{\dagger,cf}_{2,-k} \Psi^{cf}_{1,k}
\end{align}
Hence, such bilayer system could be immune to any PH invariant disorder. If we measure the PH even operators with respect to interlayer scattering, there should be an absence of $2k_f$ singularity in the static correlation function.
\begin{figure}[h]
\centering
  \includegraphics[width=0.5\linewidth]{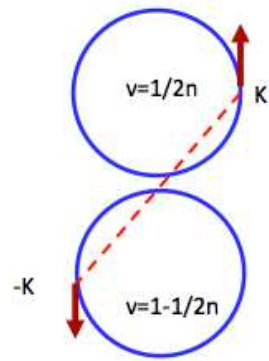}
  \caption{interlayer back-scattering in bilayer system with $\nu_1=\frac{1}{2n}$ and  $\nu_2=1-\frac{1}{2n}$ on each layer }
\end{figure}

However, if the two layers with $\nu=\frac{1}{2n}$ and  $\nu=1-\frac{1}{2n}$ are totally decoupled, the interlayer U(1) gauge symmetry is preserved. Any interlayer tunneling operator which breaks this U(1) gauge symmetry would have a vanishing expectation value. Hence, in order to measure the interlayer back-scattering susceptibility, we shall turn on weak interlayer coherence $\Delta \Psi^{\dagger,cf}_{2} \Psi^{cf}_{1}+h.c.$ which effectively turns on tunneling between composite Fermi surfaces. The two degenerate composite fermi surfaces from two layers split into enlarged/shrunk Fermi surface with Fermi momentum $k_f \pm \sqrt{\Delta }$. We assume $\sqrt{\Delta }<< k_f$ so the low energy physics is still near $k\sim k_f$ region. The two enlarged/shrunk Fermi surfaces are composed of the bonding/anti-bonding state $\Psi^{cf}_{\pm}=\Psi^{cf}_{1} \pm \Psi^{cf}_{2}$ between the two layers. During this layer coherence procedure, the PH symmetry is unbroken so the total Berry phase of the two Fermi surface remains $2\pi$. 
The PH-odd back-scattering could be expressed as,
\begin{align} 
&\Psi^{\dagger,cf}_{2,-k} \Psi^{cf}_{1,k}+h.c=\Psi^{\dagger,cf}_{+,-k} \Psi^{cf}_{+,k}-\Psi^{\dagger,cf}_{-,-k} \Psi^{cf}_{-,k}+h.c
\end{align}
One could then define the PH even interlayer tunneling operator and the absence of  $2 (k_f \pm \sqrt{\Delta})$ singularity could be a smoking gun to probe the fractional Berry phase. However, as the composite fermions come from the vortex of the physical electrons, one shall not write down a composite fermion tunneling term at UV level in the Hamiltonian. Instead, such coherence and tunneling at composite fermion level can only be realized via strong interaction\cite{alicea2009interlayer,you2017interlayer}, which makes it challenging to verify in numerical simulations. In addition, even the $2k_f$ singularity is suppressed at lowest order, there still exist some PH even multi-fermion back-scattering process as $(\Psi^{\dagger,cf}_{2,-k} \Psi^{cf}_{1,k}+h.c)(\Psi^{\dagger,cf}_{1,k'} \Psi^{cf}_{1,k'}-\Psi^{\dagger,cf}_{2,-k'} \Psi^{cf}_{2,-k'})$, so higher order singularity at $2k_f$ is still expected.

\section{competing orders in PH partner bilayer system}
\subsection{exciton condensation from interlayer pairing}
The partial filled Landau level forms a stable vortex metal at even denominator. When we take a bilayers system 
at $\nu_1=\frac{1}{2n}$ and $\nu_2=1-\frac{1}{2n}$, where each layer is the PH partner of the other,  the bilayer system could exhibit rich phase diagram with various competing orders.

For half-filled LL bilayer($n=1$), the composite Fermi surface encounters with interlayer pairing instability at short distance and the corresponding state is the known as exciton condensate where electron and hole bound together toward superfluidity\cite{eisenstein2014exciton,eisenstein2004bose}.  When half-filled LL bilayer is placing at intermediate distance, there could appear some other instabilities such as interlayer coherent composite Fermi liquid state\cite{you2017interlayer,alicea2009interlayer}, where the composite fermion acquires layer coherence and tunneling. In addition, when the composite Fermi surface has both interlayer coherence and interlayer pairing, the bilayer system is fully gapped without any symmetry breaking and the corresponding phase exhibit $Z_4$ topological order\cite{sodemann2016composite}.

To investigate the competing order in bilayer system with $\nu_1=\frac{1}{2n}$ and $\nu_2=1-\frac{1}{2n}$, we start with the situation where the layers are fallen apart with two independent composite Fermi surface,
 \begin{align} 
&\mathcal{L}=J^{cf,1}_{\mu} (a^-_{\mu}+a^+_{\mu})-\frac{1}{2n\pi}A^+ da^+ +\frac{1}{8n\pi}A_1 dA_1\nonumber\\
&J^{cf,2}_{\mu} (-a^ -_{\mu}+a^+_{\mu})-\frac{1}{2n\pi}A^- da^- +\frac{2n-1}{8n\pi} A_2 dA_2\nonumber\\
&+\frac{V(r)}{g}(d a^+)^2+\frac{V'(r)}{g'}(d  a^-)^2\nonumber\\
&a^{\pm}=\frac{a_1\pm a_2}{2},A^{\pm}=\frac{A_1\pm A_2}{2}
\label{layer}
\end{align}
Here $J^{cf}$ is the composite fermion current and $a_1,a_2$ label the internal gauge field on each layer which refer to the electron degree of freedom. $A_1,A_2$ are the external electromagnetic field on each layer.
The Maxwell term for gauge fluctuation indicates the interaction between electrons. If we assume the electron interaction has the form of Coulomb potential, the interaction for $a^+=\frac{a_1+ a_2}{2}$ is long-ranged so $V(r)=1/r$. 
Meanwhile, as the gauge flux for $a^-=\frac{a_1- a_2}{2}$  refers to an electron-hole bound state between layers, the interaction for $a^-$ is short-ranged in the form of dipole interaction $V'(r)=1/r^3$\cite{sodemann2016composite}. Therefore, the gauge fluctuation at long wave-length is,
 \begin{align} 
&\mathcal{L}_a(q)=\frac{q}{g}(a^+)^2+\frac{q^2}{g'}(a^-)^2
\end{align}
When these gauge fields are exposed to the composite Fermi surface, the coupling constant would flow as\cite{Metlitski-2015b},
 \begin{align} 
&\frac{d g}{d l}=-g^2,~\frac{d g'}{d l}=\frac{g'}{2}-(g')^2
\end{align}
Hence, the gauge fluctuation of $a^+$ is strongly suppressed by the Fermi surface while the gauge fluctuation of $a^-$ remains with a finite coupling constant $g'$.

If one reduce the layer distance, the gauge fluctuation $a^-$ introduces pairing instability between layers. Denote $V^+ / V^-$ as the intra/interlayer BCS pairing vertex,
 \begin{align} 
&\frac{d V^+}{d l}=-(V^+)^2+g',~\frac{d V^-}{d l}=-(V^-)^2-g'
\end{align}
The fluctuation for $a^-$ would suppress the intralayer pairing but enhance the interlayer pairing. Here we ignore the effect from $a^+$ as its fluctuation is already screened by the composite Fermi surface. Since the pairing gaps out the composite Fermi surface and breaks the U(1) gauge symmetry, the gauge field $a^+$ is Higgsed and the remaining gauge field $a^-$ is gapless with a survival Maxwell term.

 \begin{align} 
&\mathcal{L}_a(q)=\frac{1}{g'}(\partial a^-)^2-\frac{1}{2n\pi}A^- da^-
\end{align}
As the $4n\pi$ flux of $a^-$ is the electron-hole bound state between two layers, the gapless nature of $a^-$ indicates the system was in the exciton condensation phase where the physical electron and hole on each layer condensed to form a (111) state ($\Psi^{\dagger,e}_1\Psi^{e}_2 \neq 0$). The Maxwell term of $a^-$ refers to the Goldstone mode with respect to interlayer U(1) symmetry breaking, which could be viewed as the dual description of exciton condensate.

\subsection{$Z_{4n}$ topological order}
Apart from the bilayer pairing state, the two composite Fermi surface from each layer could also form an interlayer coherent composite Fermi liquid(ICCFL) state\cite{you2017interlayer,alicea2009interlayer}. This state spontaneously generates coherence($\phi_1=\langle \Psi^{\dagger,cf}_{2} \Psi^{cf}_{1} \rangle \neq 0$) between composite fermions(not electrons) and breaks the U(1) gauge symmetry for $a^-$.  The consequential state is still metallic, while the composite Fermi surfaces from different layer form bonding/anti-bonding states with an enlarged/shrunk composite Fermi surface.

During the phase transition toward ICCFL ,the critical boson $ \phi_1=\langle \Psi^{\dagger,cf}_{2} \Psi^{cf}_{1} \rangle$ is massless around the critical region and the Landau damping effect of the critical boson becomes more relevant. At the critical region, the strong fluctuation of critical boson $\frac{q^2}{g''}|\phi_1|^2$ could enhance the interlayer pairing channel. 
 \begin{align} 
&\frac{d V_s^-}{d l}=-(V^-)^2-g'-g'',\frac{d V_a^-}{d l}=-(V^-)^2-g'+g''
\end{align}

\begin{figure}[h]
\centering
  \includegraphics[width=0.7\linewidth]{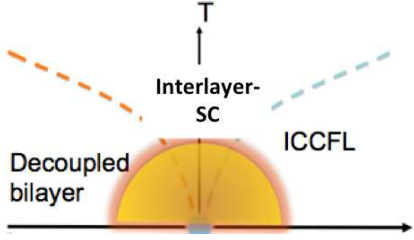}
  \caption{Phase diagram for the ICCFL transition. The SC dome covers the quantum critical region for ICCFL transition. There exist a region where $ \phi_1= \langle \Psi^{\dagger,cf}_{2} \Psi^{cf}_{1} \rangle \neq 0$ and $\phi_2=\langle \Psi^{cf}_{2} \Psi^{cf}_{1} \rangle \neq 0$.}
  \label{phased}
\end{figure}

$V_s^-$ and $V_a^-$ refers to the interlayer pairing vertex which is symmetric/asymmetric among layers. Since both gauge fluctuation of $a^-$ and the critical boson $\phi_1$ enhance the symmetric interlayer pairing state, the critical regime near the ICCFL transition could be covered by the superconductivity(SC) dome with respect to interlayer pairing, as Fig \ref{phased}. Thereby, there appears a regime where a SC dome overlaps with interlayer coherence where $ \phi_1= \langle \Psi^{\dagger,cf}_{2} \Psi^{cf}_{1} \rangle \neq 0$ and $\phi_2=\langle \Psi^{cf}_{2} \Psi^{cf}_{1} \rangle \neq 0$. The phase is fully gapped with no explicit symmetry breaking. Indeed, such phase carries intrinsic topological order. 

To explore the topological structure of such phase, 
we can write down the field theory of $\phi_1$ for ICCFL transition,
\begin{align} 
&\mathcal{L}= |(\partial_{\mu}+2a^-_{\mu})\phi_1|^2-\frac{1}{2 \pi n}a^-_{\mu} \partial A^-+a|\phi_1|^2+b|\phi_1|^4
\end{align}
As the transition concur with interlayer pairing, the system goes into an exciton condensate phase near the criticality and the $4n \pi$ monopole of $a^-$ is the exciton order parameter. Due to the gauge field $2a^-$ carried by $\phi_1$, in the background of exciton condensate, $\phi_1$  is equivalent to the $8n \pi$ vortex of the exciton order parameter. The condensation of $\phi_1$ is then akin to condensation of $4n$ bundles of $2\pi$ vortices for the exciton order parameter.  As a result, the condensation of $\phi_1$, which Higgesd $a^-$, would kill the Goldstone mode for interlayer U(1) and restore the symmetry. Meanwhile, as the condensation of $\phi_1$ is equivalent to  the condensation of $4n$ bundle of $2\pi$ vortices, its resultant state shall exhibit $Z_{4n}$ gauge theory so the phase contains $Z_{4n}$ topological order.

Recent numerical simulation\cite{zhuzheng} had shown that the bilayer Half-filled Landau levels at intermediate layer distance could emerge a fully gapped phase between the exciton condensate and the decoupled composite Fermi liquid phase. Such phase contains a gapped spectrum and flat Berry curvature distribution in both charge U(1) and interlayer U(1) sector. This suggests the ground state does not break either charge U(1) or interlayer U(1) symmetry. 

Here we point out a possibility that the exotic intermediate phase revealed by the numerical result in \cite{zhuzheng} is the $Z_{4}$ topological order phase where both $ \phi_1= \langle \Psi^{\dagger,cf}_{2} \Psi^{cf}_{1} \rangle \neq 0$ and $\phi_2=\langle \Psi^{cf}_{2} \Psi^{cf}_{1} \rangle \neq 0$. 
Although the condensation of $\phi_1$ and $\phi_2$ requires two parameters, the critical boson $\phi_1$ which drives the coherence between two composite fermi surface could enhance the interlayer pairing $\phi_2=\langle \Psi^{cf}_{2} \Psi^{cf}_{1}\rangle $. Such quantum criticality enhanced pairing mechanism drives a coexistence phase where the two composite Fermi surfaces have both interlayer pairing and interlayer coherence. This generates a symmetry invariant fully gapped phase with topological order.

\section{Conclusion and remark}
At this stage, we revisit the partial filled Landau level with even filling fraction. We demonstrate that the IR theory for $\nu=\frac{1}{2n}$ filled Landau level contains a composite Fermi surface formed by the vortex of the electron. Such vortex metal is a non-Fermi liquid where the quasiparticle near the Fermi surface is decoherent by the gauge bosons. The composite Fermi surface for $\nu=\frac{1}{2n}$ contains $-\pi/n$ Berry phase, which is a consequence of LL projection and guiding center algebra. Such Berry phase evokes an anomalous velocity to the composite Fermion's equation of motion. To measure the Berry phase effect, we look into the bilayer partial filled Landau level at $\nu_1=\frac{1}{2n}$ and $\nu_2=1-\frac{1}{2n}$. Such PH symmetric bilayer system contains two composite Fermi surface with the same Fermi wave vector, each carries a Berry phase $-\pi/n$ and  $2\pi+\pi/n$. The overall Berry phase of the bilayer vortex metal is protected by PH symmetry. The back-scattering between two layers is suppressed so the system is immune to any PH symmetry invariant disorder. We also investigate the bilayer instability in PH partner bilayer, and introduce a new $Z_{4n}$ topological order phase which partly agrees with recent numerical result\cite{zhuzheng}.

\begin{acknowledgments}
We are grateful to Chong Wang, Olexei Motrunich, Barry Bradlyn, Mike Zaletel, Biao Lian, Sri Raghu, Eduardo Fradkin, Yi-Zhuang You and Donna Sheng for insightful comments and discussions. Y-Y is supported by the PCTS Fellowship at Princeton University. Y-Y acknowledge SCGP where the initial idea of this draft was created.

\end{acknowledgments}

\appendix
\section{Kramers degeneracy for PH symmetry}
\begin{widetext}
In this appendix, we would show that PH symmetry operation in bilayer system with $\nu=\frac{1}{2n}$ and  $\nu=1-\frac{1}{2n}$ is projective and contains Kramers degeneracy.

One could write the wave function of the partial filled LL bilayer as,
\begin{align} 
\Psi=\prod_{i}^{N_1} c_i^{\dagger} c^{\dagger}_j  ...\prod^{N_2}_{i'} d^{\dagger}_{i'} d^{\dagger}_{j'}.. |0\rangle
\end{align}
$c$ and $d$ are the electron creation operator on each layer with filling $\nu_1=\frac{1}{2n}$ and  $\nu_2=1-\frac{1}{2n}$. $|0\rangle$ is the empty LL for 2 layers. The total number of the fermion $N_1+N_2$ on two layers is half of the total orbitals $2N$.

Now applying the PH symmetry $\mathcal{CT}$. While the PH operator takes the electron creation operator to annihilation operator, it also changes the vacuum from an empty LL toward a filled LL.
\begin{align} 
&\mathcal{CT} \Psi=\prod^{N_2}_{i} c_i c_j  ...\prod^{N_1}_{i'} d_{i'} d_{j'} |1\rangle\nonumber\\
 &=\prod^{N_2}_{i} c_i c_j  ...\prod^{N_1}_{i'} d_{i'} d_{j'} \prod^{N}_{i} c^{\dagger}_i c^{\dagger}_j  ...\prod^{N}_{i} d^{\dagger}_{i'} d^{\dagger}_{j'}  |0\rangle
\end{align}
$|1\rangle$ is the filled 0th LL for 2 layers.
Acting the PH symmetry twice, one obtains
\begin{align} 
&(\mathcal{CT})^2 ~\Psi=\prod^{N_1}_{i} c_i^{\dagger} c^{\dagger}_j  ...\prod^{N_2}_{i'} d^{\dagger}_{i'} d^{\dagger}_{j'}..\prod^{N}_{i} c_i c_j  ...\prod^{N}_{i} d_{i'} d_{j'}.. \prod^{N}_{i} c^{\dagger}_i c^{\dagger}_j  ...\prod^{N}_{i} d^{\dagger}_{i'} d^{\dagger}_{j'}  |0\rangle  \nonumber\\
&=\prod^{N_1}_{i} c_i^{\dagger} c^{\dagger}_j  ...\prod^{N_2}_{i'} d^{\dagger}_{i'} d^{\dagger}_{j'}.. \prod^{N}_{i} c_i c^{\dagger}_i  c_j c^{\dagger}_j   ...\prod^{N}_{i'} d_{i'} d^{\dagger}_{i'}  d_{j'} d^{\dagger}_{j'}  (-1)^{N(2N-1)} |0\rangle \nonumber\\
&= (-1)^{N(2N-1)} \prod^{N_1}_{i} c_i^{\dagger} c^{\dagger}_j  ...\prod^{N_2}_{i'} d^{\dagger}_{i'} d^{\dagger}_{j'}..  |0\rangle = (-1)^{N}\Psi
\end{align}
Acting the PH symmetry twice gives a phase factor $(-1)^{N}$. 
\end{widetext}
\providecommand{\noopsort}[1]{}\providecommand{\singleletter}[1]{#1}%

\end{document}